\documentclass[twocolumn,aps,pra]{revtex4}

\usepackage{graphicx}
\usepackage{dcolumn}
\usepackage{bm}
\DeclareGraphicsExtensions{.jpg,.pdf}
\begin{document}

\title{Regular Spectra and Universal Directionality of Emitted Radiation from a Quadrupolar Deformed Microcavity}

\author{Jeong-Bo Shim}
\author{Hai-Woong Lee}
\email{hwlee@laputa.kaist.ac.kr}
\affiliation{Department of
Physics, Korea Advanced Institute of Science and Technology,
Daejeon, 305-701, Korea}

\author{Sang-Bum Lee}
\author{Juhee Yang}
\author{Songki Moon}
\author{Jai-Hyung Lee}
\author{Kyungwon An}
\affiliation{School of Physics, Seoul National University,
Seoul, 151-742, Korea}

\author{Sang Wook Kim}
\affiliation{Department of Physics Education, Pusan National
University, Busan 609-735, Korea}

\date{\today}

\begin{abstract}
We have investigated quasi-eigenmodes of a quadrupolar deformed
microcavity by extensive numerical calculations. The spectral
structure is found to be quite regular, which can be explained 
on the basis of the fact that the microcavity is an open system. The far-field
emission directions of the modes show unexpected similarity
irrespective of their distinct shapes in phase space. This
universal directionality is ascribed to the influence from the
geometry of the unstable manifolds in the corresponding ray
dynamics.
\end{abstract}

\maketitle

\section{Introduction}

Symmetrical (cylindrical or spherical) microcavities have
attracted much attention in the past as laser resonators, due to
their compact and simple geometry, easiness to fabricate, and
ultra-high Q values, which is attributed to formation of the
so-called whispering gallery modes (WGM's) \cite{yamamoto}. The
isotropic radiation pattern of light emitted from the symmetrical
microcavity, however, reduces their practical usefulness. One way
to overcome this problem, as has been proposed by N\"ockel et al.,
is to deform the shape to construct asymmetric resonant cavities
(ARC's) \cite{nockel,stone1}. Directional emission has indeed been observed
experimentally from asymmetrically deformed microlasers made of
semiconductors \cite{gmachl,rex}, dye jets \cite{lee} and polymers \cite{harald}.

Fundamental optical properties of ARC's can be partly understood
in terms of the ray dynamics. From the classical dynamical
viewpoint, it belongs to a class of systems having mixed-type phase
space, i.e., a mixture of both regular and irregular trajectories
\cite{swkim,percival,bohigas}. When deformation is very
small, the phase space is dominated by regular rays of whispering
gallery type and these rays escape from the cavity only by
evanescent leakage (tunnelling). At larger deformations, however,
emission properties are mainly determined by chaotic rays that
diffuse stochastically and refract out of the resonator when the
angle of incidence $\chi$ reaches the critical angle $\chi_c$ ,i.e.,
$\sin \chi_c = 1/n$, where $n$ is the refractive index of the
resonator. This simple ray dynamics model predicts that at large
deformations the escape occurs primarily into the tangential
direction near the points of maximum curvature \cite{nockel,nockel2}.

Although the model captures the essence of the directional
property of emitted radiation from ARC's, there may exist other
complications. For example, a phenomenon referred to as dynamical
eclipsing \cite{Chang} occurs when stable islands occupy the phase
space region at which the escape would have taken place without them,
resulting in strong suppression of emission intensity in
directions predicted by the ray model and appearance of split
peaks in nearby directions. There are evidences that stable or
unstable periodic orbits strongly influence the emission pattern
of ARC. The emission originating from bow-tie modes \cite{gmachl,tureci}
, whose corresponding classical orbit is a stable
island, and the so-called scarred modes of hexagonal \cite{lee} 
and triangular \cite{rex} unstable periodic
orbits, has been observed. So far the direction of the far-field
emission obtained in experiments has been a commonly used tool to
investigate the characteristics of the lasing mode of ARC, since it
has been assumed that the shape of the wavefunction inside the cavity
strongly influences or sometimes even completely governs the
far-field emission pattern. It has been reported, however, that,
due to the correction with Fresnel coefficients, the emission
direction of the mode with a rather smaller $Q$ value can be
considerably deviated from an expectation based upon the above simple
assumption \cite{rex}. Deviations from the prediction of
the simple ray dynamics model also occur when the nonlinear effect
of a medium is involved \cite{harayama}.

The simple ray dynamics model has recently been refined by
Schwefel et al. \cite{harald}. They found that highly directional
emission patterns from ARC's persist well beyond the deformation
limit predicted from the ray model \cite{nockel}, and developed a
more accurate model which emphasizes the importance of unstable
manifolds of short periodic orbits. However, they did not show the
direct evidence of the crucial role played by unstable manifolds in
connetion with their structure embedded in quantum wavefunctions.
 A similar analysis has been reported by Lee
et al. in a stadium shaped microcavity, although the authors did not
precisely mention the role of the unstable manifolds \cite{paichai}.

The main purpose of this paper is to investigate optical
properties of a quadrupolar micro-resonator with low refractive
index ($n = 1.361$) by extensive numerical calculations. The reason
why we choose the specific refractive index is that this work is
initiated from our previous experiment of the dye-jet microcavity
laser \cite{lee}. We would like to deliver two main results:
firstly the energy level distribution shows deviations from an
expectation from the usual random matrix theory (RMT), and
secondly the far-filed emission direction exhibits a nontrivial
universal feature.

To a certain degree, optical properties of ARC's can be understood
by using an analogy with chaotic dynamical properties of the
billiard system which has been extensively studied in the context
of nonlinear dynamics. The quantum mechanical manifestation of
classical chaos, referred to as quantum chaos, has generally
revealed that there is a close relation between the classical
nonlinear dynamics (whether it is regular or chaotic) and quantum
spectral statistics (whether the energy level spacing distribution
is described by Poissonian or Wigner) \cite{haake}. Based on this
observation, one might expect that there also exists close
correlation between the ray dynamics and the statistical
properties of the eigemodes of ARC's. Specifically, energy levels
may be regularly distributed at small deformations, where the ray
dynamics is predominantly regular, while they may look quite
irregular at large deformations where chaotic rays dominate. We
will show, however, that this simple expectation does not exactly
hold true. There is an essential difference between ARC and the
billiard system: The former is an open system while the latter is
a closed one. The openness of ARC raises a nontrivial question:
Where is the emitted output directed? This is completely
meaningless in a closed system. Although the analogy between ARC
and the billiard system is a useful starting point because the
billiard has been thoroughly studied in the past, one cannot push
the analogy too far.

This paper is organized in the following way: In Section II, we
briefly explain both the ray dynamics and the wave properties of a
quadrupolar billiard, and a quadrupolar deformed microcavity
(QDM). In Section~III, we show numerical results for spectra and
far-field emission directionality of a QDM and discuss the
regularity of the spectra and the universal
directionality we obtained from the numerical computations.
 Finally in Section IV, we present a conclusion.

\section{A Quadrupolar Deformed Microcavity}

When a liquid jet with circular cross section is deformed, its
cross section can in general be described by the following
multipole expansion \cite{lamb}
    \begin{eqnarray}
    \label{eq:expansion}
r(\phi ) = \frac{1}{2}a_0  + \sum\limits_{j = 1}^\infty  {a_j \cos
j\phi }  + \sum\limits_{j = 1}^\infty  {b_j \sin j\phi },
    \end{eqnarray}
where $r$ is the distance from the center and $\phi$ the azimuthal
angle. The lowest order even-symmetric quadrupolar contribution
becomes dominant when the perturbation inducing deformation is
small. The cross section of the dye jet can then be described by
 \begin{eqnarray}
    \label{eq:quadrupole}
    r(\phi)=\frac{r_0}{\sqrt {1+ \varepsilon ^2 /2}}(1 + \varepsilon \cos
    2\phi),
 \end{eqnarray}
where $r_0$ and $\varepsilon$ are respectively the radius of the
undeformed (circular) cavity and the deformation parameter. The
denominator in Eq.~(\ref{eq:quadrupole}) guarantees that the cross
sectional area is conserved irrespective of $\varepsilon$, which is
 consistant with the experimental condition of Ref. \cite{lee}. In this
paper we focus on such a specific ARC, i.e., a QDM.

\subsection{A quadrupolar billiard: closed system}

\begin{figure}
\centering
\includegraphics[clip=true,width=7.5cm]{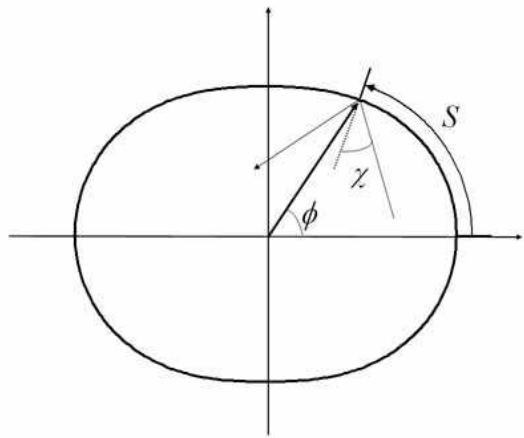}
\caption{A schematic view of Birkhoff coordinates: They consist of
the distance along the boundary $S(\phi)$ and the reflection
angle, $\chi$. To make them canonically conjugate, $\sin\chi$ is
used instead of $\chi$.}\label{fig:birkhoff}
\end{figure}

\begin{figure*}
\centering
\includegraphics[width=15cm]{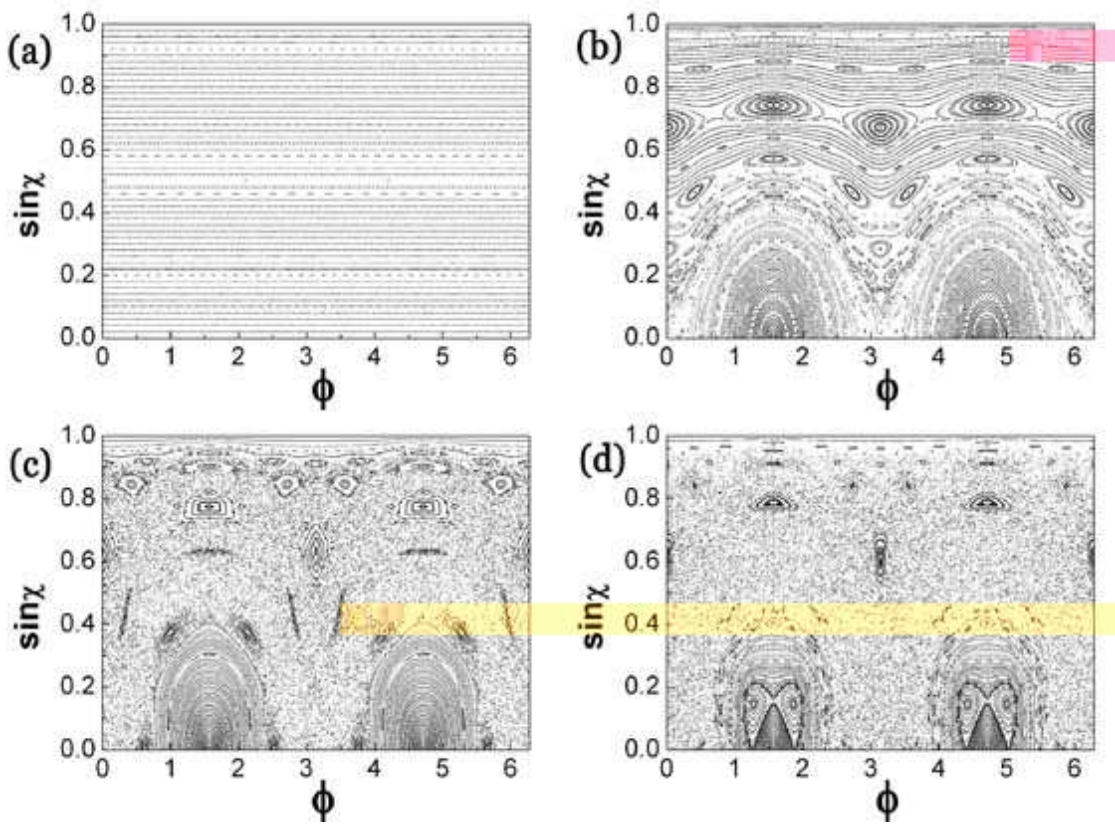}
\caption{The Poincares surfaces of section of a quadrupolar
billiard for (a) $\varepsilon = 0$, (b) $\varepsilon = 0.05$, (c)
$\varepsilon = 0.10$, and (d) $\varepsilon = 0.12$.}\label{fig:sos}
\end{figure*}

At first, we consider the ray dynamics of a QDM with hard walls,
namely a quadurpolar billiard. The Birkhoff coordinate system is quite
useful to study the ray dynamics in a billiard. It consists of two
variables $S$ (or equivalently $\phi$) and $\sin\chi$, which
respectively represent the arc length of the position at which the
ray strikes the boundary and the sine of the incident (or
reflected) angle at the boundary \cite{birk}
(Fig.~\ref{fig:birkhoff}). In Fig.~\ref{fig:sos} we show the
Poincare surfaces of section (PSOS) of the dynamics in the
quadrupolar billiard by using Birkhoff coordinates. It exhibits a
typical Kolmogorov-Arnold-Moser scenario \cite{lichtenberg}, i.e.
the evolution toward stochastic behavior as the deformation
parameter $\varepsilon$ is increased.

\begin{figure}
\centering
\includegraphics[width=8.5cm]{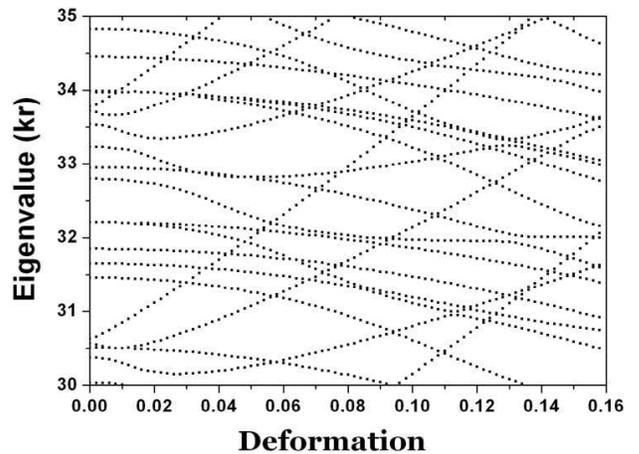}
\caption{Evolution of the eigenvalues as a function of $\varepsilon$.
It shows a complicated structure.}\label{fig:evol_closed}
\end{figure}

The wave nature of a quadrupolar billiard can be described by the
following Helmholtz equation with Dirichlet boundary condition
\begin{eqnarray}
\label{helmholtz} \nabla ^2 \psi(r)  + k^2 \psi(r)  = 0,
\end{eqnarray}
where $k$ is the wave number. For a circular billiard,
 the solution of Eq.~\ref{helmholtz} takes the form 
$J_m(r)\exp(im\phi)$, where $J_m$ is a Bessel function. All
the eigenvalues $k_{lm}$ are analytically determined and form a
well organized structure since they are just the $l$th zeros of
the $m$th order Bessel functions. When a circular billiard is
deformed, however, the distribution of eigenvalues starts to
deviate from the regular structure. For large enough deformations,
in which case the corresponding ray dynamics is predominantly
chaotic, the well established RMT plays an important role.
Although the nearest neighbor level spacing distribution of
eigenvalues is described by the universal Wigner function, the
eigenvalues themselves show quite a complex structure due to a lot
of avoided crossings. Fig.~\ref{fig:evol_closed} shows the
evolution of eigenvalues ($kr$) as a function of the deformation
parameter.

\subsection{A quadrupolar deformed microcavity: open system}

A QDM is an open system in the sense that the ray with the incident angle
larger than a critical angle is allowed to escape from the cavity.
It does not make any influence on the ray dynamics itself inside
the cavity, but reduces available phase space: e.g. for $n=1.361$
the phase space area below $\sin\chi_c = 1/n \approx 0.73$ can be
ignored.

The wave properties of an open system somewhat differ from those
of a closed one. The eigenvalues of Eq.~(\ref{helmholtz}) now
take complex values. Their imaginary part is associated with
the decay rate of the mode. To quantify the decay rate, a cavity
 quality factor $Q$ defined as $-2 \times k_r/k_i$, where $k_r$ and $k_i$ are respectively
real and imaginary parts of the eigenvalue, is commonly used . The
real part of the eigenvalue is represented by the size parameter
defined as $nk_rr_0$, which is abbreviated as $nkr$ if there is no
confusion.

As far as a circular microcavity is concerned,
Eq.~(\ref{helmholtz}) is still separable so that two good quantum
numbers, radial ($l$) and angular ones ($m$), can be well defined.
The eigenvalues $k_{lm}$ are analytically given by matching
between Bessel and Neumann functions. Similarly to a circular
billiard, one can see the well-defined free spectral range (FSR)
among the modes with the identical radial quantum number $l$.

\begin{figure}
\centering
\includegraphics[clip=true,width=8.5cm]{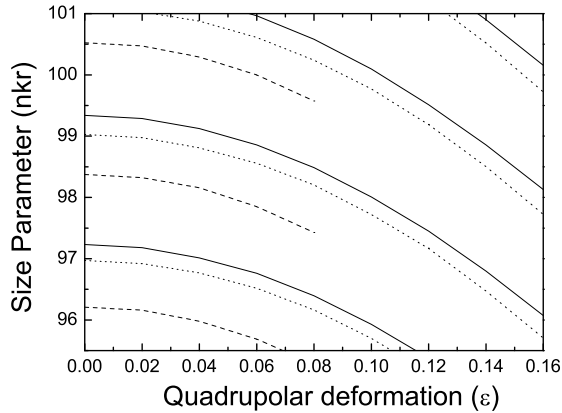}
\caption{Evolution of the real part of the eigenvalues of the mode
with $l=1$ (the dotted line), $l=2$ (the solid line), and
$l=3$ (the dashed line) of a QDM as a function of
$\varepsilon$.}\label{fig:redshift}
\end{figure}

Figure \ref{fig:redshift} presents a continuous variation of the
real part of eigenvalues as the deformation is increased.
Surprisingly, the eigenvalues vary quite regularly in comparison
with the complicated structures in Fig.~\ref{fig:evol_closed}.
This regular structure remains, although the deformation becomes
large enough to generate global chaos in classical dynamics. The
monotonic decrease of $nkr$ implies the red shift of spectral
lines. In case of WGM it can be explained by considering an elongated
optical path due to deformation. In order to explain the
regularity of spectrum, we need to investigate more carefully the
nature of the modes. In the following section their phase space
distribution will be discussed based mainly upon the Husimi plot.

When the system is open, a question naturally arises: Into which
direction is the output field emitted? In a circular cavity,
the emission direction should be isotropic due to the circular
symmetry, while it is highly non-trivial to address it for ARC's.
In the viewpoint of the ray dynamics, the emission direction is
determined from the position ($\phi$,$\sin\chi$) at which the incident
angle of the ray becomes equal to the critical angle. In wave
mechanics, however, one should find quasi-eigenmodes, which are an
open system analogue to eigenmodes of a closed system, by direct
numerical calculation. In this work we use the boundary element
method to compute quasi-eigenmodes \cite{kagami,wiersig}.

\section{Regular spectrum and universal emission directionality}

In this section we discuss the origin of the regularity of the
spectrum and the universal emission directionality. The Husimi
distribution function is a very powerful tool to study these issues.

\subsection{The Husimi plot of the eigenmode of the deformed microcavity}

The Husimi distribution function can be regarded as a mathematical
description of the phase-space distribution in which the
minimum-uncertainty Gaussian wave packet is used as a basis \cite{hwlee}:
\begin{eqnarray}
\label{husimiconv} H_\psi (q_0,k_0) =\left| \int {\psi (q)\;\; \xi
(q-q_0,k_0)} \;dq \right|^2,
\end{eqnarray}
where
\begin{eqnarray}
\label{packet} \xi (q-q_0,k_0)=\exp \left[-\frac{{(q - q_0 )^2
}}{{2\sigma }} + ik_0 (q - q_0 )\right].
\end{eqnarray}
In order to obtain the Husimi plot in Birkhoff coordinates, it
is necessary to modify Eqs. (\ref{husimiconv}) and (\ref{packet}).
Taking it into account that $\phi$ is $2\pi$ periodic, the basis
function $\xi$ can be rewritten as
\begin{eqnarray}
\label{xinew} \nonumber \xi (\phi-\phi_0,\sin \chi) =
\sum\limits_l \exp \left[ - \frac{{(\phi  + 2\pi l - \phi _0 )^2
}}{{2\sigma }}\right.\\ \left. + ik\sin \chi (\phi  - \phi _0 )
\right],
\end{eqnarray}
and the integration in Eq.~(\ref{husimiconv}) should be taken from
0 to $2\pi$ \cite{Crespi,martina}. For our calculation we set the
width of the gaussian as $\sigma = \sqrt{2} / k_0$.

\begin{figure*}
\centering
\includegraphics[clip=true,width=15cm]{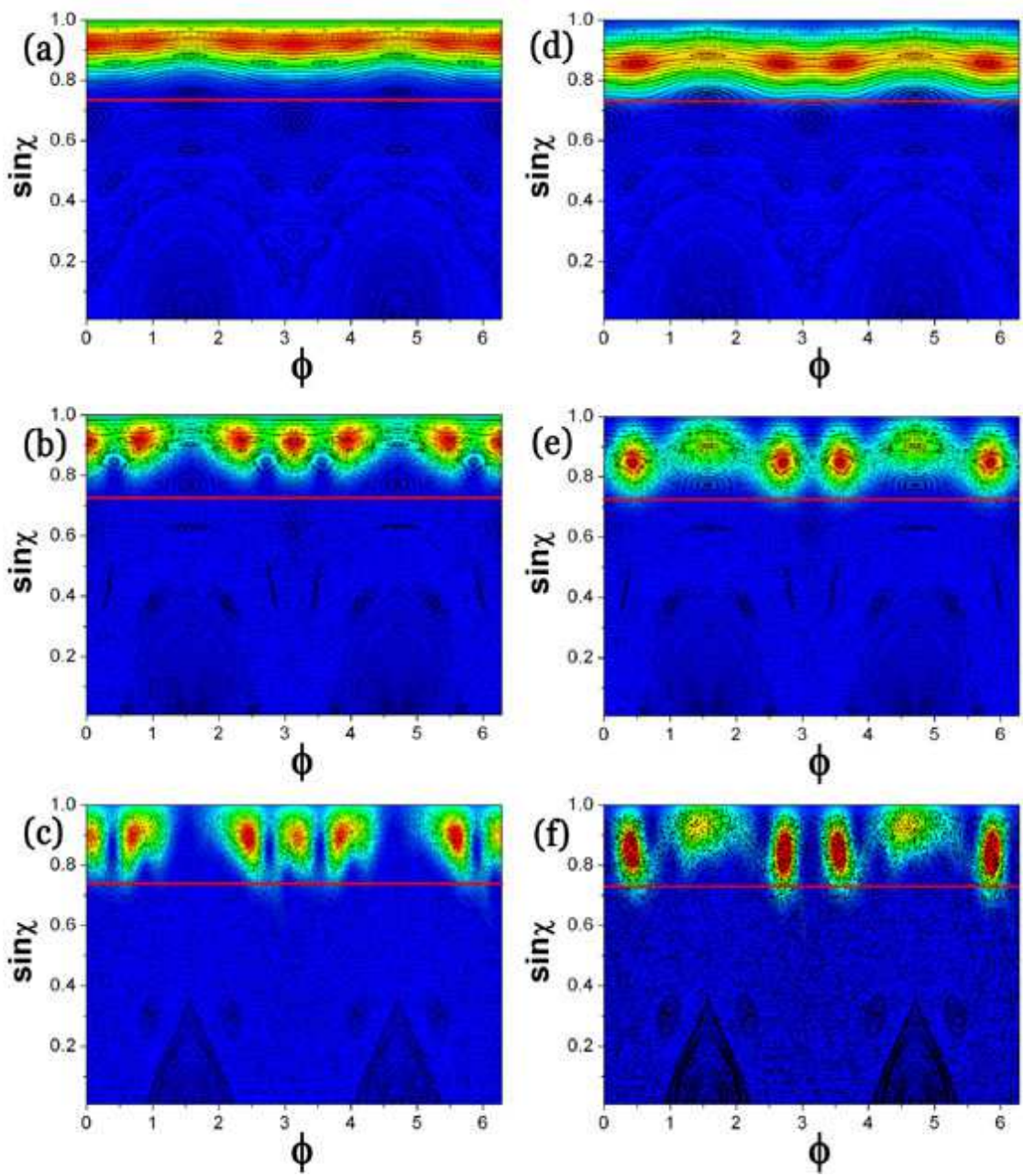}
\caption{Husimi plots of a QDM with (a)$\varepsilon = 0.05$, (b)
$0.1$ and (c) $0.16$ for the $l=1$ modes, and with (d)
$\varepsilon = 0.05$, (d) $0.1$ and (f) $0.16$ for the $l=2$ modes.
 The corresponding complex size parameters are (a)
$96.64-i0.003$, (b) $96.89-i0.004$, (c) $95.87-i0.004$, (d)
$96.10-i0.005$, (e) $96.25-i0.014$, and (f) $96.61-i0.029$.
The critical line defined by $sin \chi_c = 1/n$ is indicated with the 
red horizontal lines.}\label{fig:evolution}
\end{figure*}

\begin{figure*}
\centering
\includegraphics[clip=true,width=15cm]{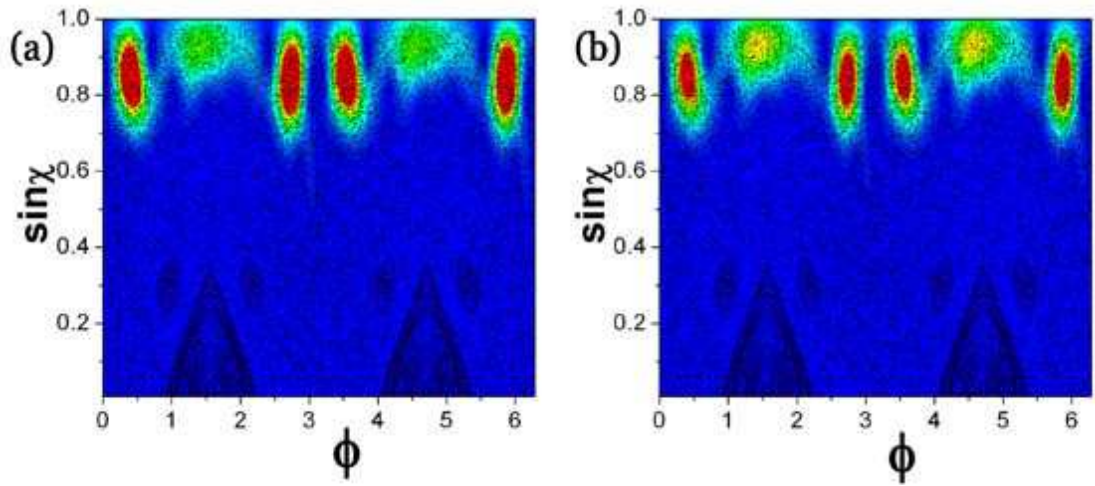}
\caption{Husimi plots of the $l=2$ mode at (a) $nkr=96.073$
and (b) $nkr=98.12$ for $\varepsilon =0.16$. These two modes are
separated by two units of FSR.}\label{fig:husimiad}
\end{figure*}

In Fig.~\ref{fig:evolution} we present Husimi plots of several
quasi-eigenmodes obtained numerically. The $l=1$ modes in a
circular cavity evolves from a horizontally  flat shape
 to the distribution localized
somewhere in the phase space in Fig.~\ref{fig:evolution}(c).
A similar evolution is also observed in the $l=2$ modes except
that it is localized on the period-6 orbit which loses its
stability around $\varepsilon \sim 0.12$. The $l=2$ modes then
forms a hexagonal shaped scarred mode extensively investigated in
our previous work \cite{lee}. All the modes identified
with $l=2$ in Fig. \ref{fig:redshift} appear to have the similar
shapes in Husimi plots as shown in Fig. \ref{fig:husimiad}. It is
more difficult to identify the corresponding classical orbit of
the $l=1$ modes. It seems to correspond to the period-6 orbit
based upon Fig. \ref{fig:evolution}(c), since it has six enhanced
localized probability maxima in the phase space. However, their
locations along the vertical axis ($\sin\chi$) fit themselves to
those of the period-8 orbit. Interestingly, this mode exactly
corresponds to the period-8 orbit for larger $nkr$. We believe that
the $l=1$ mode is experiencing the transition from the period-6
to the period-8 orbit. More details will be discussed elsewhere.

\subsection{Regular spectra}

In closed systems, chaotic states which are irregularly
distributed over available phase space are frequently observed, so
that they strongly interact with and repel each other to generate
avoided crossings and form complicated spectral structures as
shown in Fig.~\ref{fig:evol_closed}. In contrast, quasi-eigenmodes
in open systems hardly form chaotic states. The reason is that the
chaotic modes have too small $Q$ values to form modes in the practical
sense, since they decay in a short time due to refractive escape. In
Fig.~\ref{fig:redshift} the $l=3$ mode seems to disappear at
$\varepsilon \sim 0.08$, which implies that it is difficult to find
the mode due to a very small $Q$, since, for a given $nkr$, the larger
$l$ is the smaller $Q$ is. The mode with even larger $l$ ($>3$) may
exist in Fig.~\ref{fig:evol_closed} but is practically meaningless,
because $Q$ is extremely low. One may more frequently find chaotic
modes for larger $nkr$, because $Q$ is exponentially enhanced as
$nkr$ is increased. It leads us to the conclusion that in open
systems chaotic quasi-eigenmodes are rarely observed unless $nkr$
is large enough. 

It is no wonder that non-chaotic modes show deviations from
the prediction based upon RMT, since they are localized somewhere in
phase space. It is unlikely that for a given $nkr$ the $l=1$ mode
 has a similar shape with the $l=2$ mode. In the
 semiclassical theory \cite{nockelth} the larger $l$ the
bigger the incident angle of the ray. It implies that the modes with
different $l$'s form distinct shapes in phase space. It is clearly
shown in Figs.~\ref{fig:evolution}(c) and (f) that those two modes
hardly overlap with each other. As $\varepsilon$ is increased, they
independently evolve with their own shape maintained in phase
space. That explains the regular spectral structure observed in
Fig.~\ref{fig:redshift}.

\subsection{Universal emission directionality}

\begin{figure}
\centering
\includegraphics[clip=true,width=7.5cm]{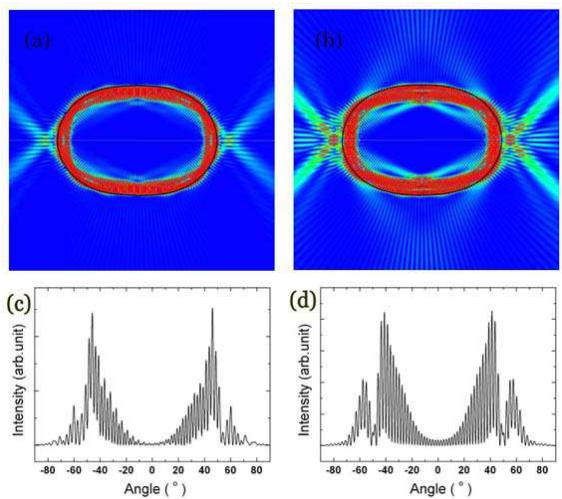}
\caption{Probability distributions of the wavefunctions of the
quasi-eigenmodes in real space for (a) $nkr=97.72$ ($l=1$) and (b)
$nkr=98.12$ ($l=2$). The far-field emission intensities are shown
for (c) the mode corresponding to (a) and (d) the mode
corresponding to (b).}\label{fig:config}
\end{figure}

The quasi-eigenmodes of an open system contain evanescent waves
outside the cavity, which cause decay of the modes in time.
Figures \ref{fig:config}(a) and (b) show that some of the waves leak
out, which cannot be seen in the Husimi plot. The output
directionality is in practice the most important characteristics
that we seek from ARC's. It is more definitely illustrated by the
far-field emission intensity as a function of the output direction
as shown in Figs. \ref{fig:config}(c) and (d). Surprisingly, these
two figures demonstrate an almost identical pattern whose maximum
is located around $40^\circ$. It should be emphasized that the
corresponding Husimi plots of these two modes differ from each
other as shown in Figs. \ref{fig:evolution}(c) and (f). Since it
has been believed that the emission direction is determined from
the internal shape of the quasi-eigenmode, it is necessary to
explain this unexpected observation.

\begin{figure*}
\centering
\includegraphics[clip=true,width=15cm]{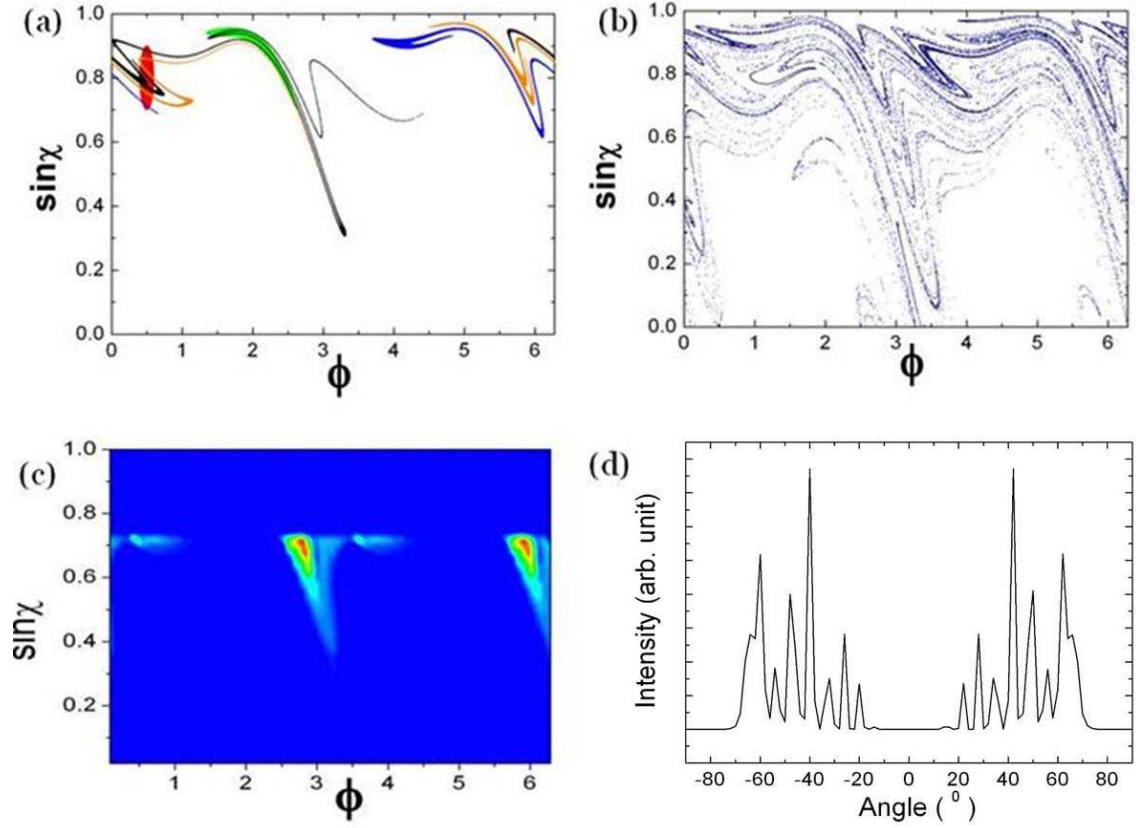}
\caption{Time evolution of the rays starting from a
set of $10^4$ initial conditions localized on the upper left side
of the phase space and far-feld emission intensity.
(a) The rays after a few steps. 
(b) The rays after 20 collisions on the boundary(the complicated structure
of unstable manifold manifest themselves). (c) The accumulated intensity of the escaped rays
weighted by Fresnel refractive coefficients.
(d) The far-field emission intensity obtained from the ray
dynamics.}\label{fig:manifold}
\end{figure*}

The clue for the origin of the identical far-field directionality
irrespective of the shapes of a quasi-eigenmode comes out of the
fact that the far-field pattern obtained from ray dynamics shown in
Fig. \ref{fig:manifold}(d) is
not much different from those from the wave calculations.
 In the ray calculations, the ray incident with the angle smaller
 than the critical angle is
reflected in or emitted out with the probability determined by
Fresnel coefficients.

In a long time the ray dynamics in the chaotic region is governed by
the so-called unstable manifolds \cite{lichtenberg}. They are
easily seen by launching a set of localized initial conditions of
rays in quadrupolar billiard and following the dynamics for a few
tens of collisions [see Figs.~\ref{fig:manifold}(a) and (b)]. As
time goes to the infinity, the ray dynamics in a closed system
will finally become ergodic so that the rays are uniformly
distributed over phase space. As far as an open system is
concerned, however, rays escape from a cavity before reaching
uniform distribution. Therefore, the available phase space is limited,
and the dynamics in the restricted space is described by the
motion along a few dominant unstable manifolds. As a result, the
unstable manifolds dominate the long (but not extremely long) time
dynamics in ARC's, and consequently the output directionality. In Fig.
\ref{fig:manifold}(c), we plot the accumulated intensity in phase space,
 obtained by recording the Birkhoff coordinates at which each ray escapes and
totaling the number of the escaped rays weighted by Fresnel refractive
coefficients at each phase space point ($\phi$,$sin \chi$). The intensitty
pattern resembles the structure of the unstable manifolds shown
in Fig. 8(b). In Fig.8(d) we show the far field emission intensity obtained
from the ray dynamics calculations, which is seen to be consistent with 
the results of the wave calculations shown in Figs. 7(c) and 7(d). 

\begin{figure*}
\centering
\includegraphics[clip=true,width=15cm]{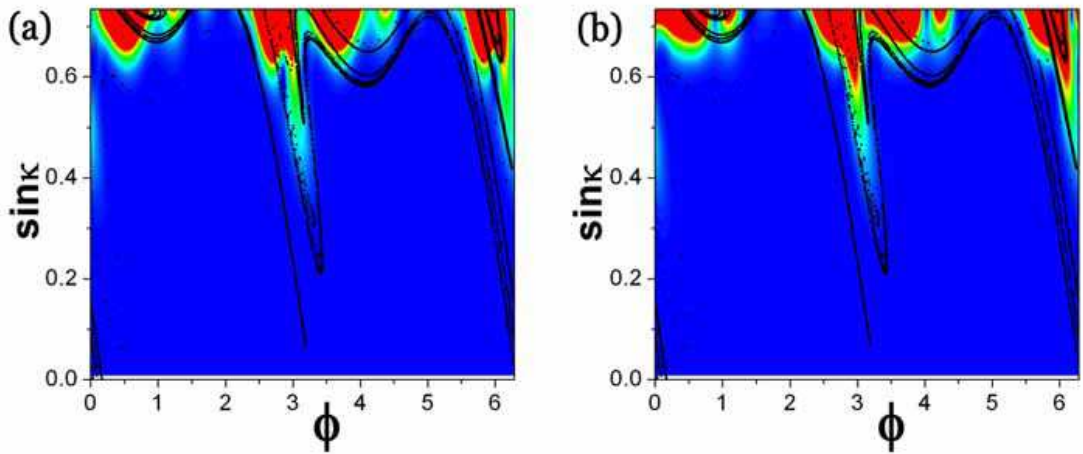}
\caption{An expanded view of the Husimi plots of (a)Fig.~\ref{fig:evolution}(c)
 and (b) Fig.~\ref{fig:evolution}(f) for the region below the critical angle. 
The unstable manifolds are shown in the background with black curves.} 
\label{fig:husimiman}
\end{figure*}

The unstable manifolds can also be found through a careful examination of
Husimi plots, as Figs.~\ref{fig:husimiman}(a) and ~\ref{fig:husimiman}(b)
 indicate.
Such a fine structure could not be seen in the original Husimi plots of Figs.~\ref{fig:evolution}, because it is associated with a tiny
probability. No matter how small the probability associated with
the unstable manifold is, it significantly overlaps the region
below the critical angle. It means that this fine structure mainly
contributes to the output directionality regardless of the tiny
probability, unless the $Q$ value is not too small. Note that $Q
\sim 10^4$ for both the $l=1$ and $l=2$ modes. We may now conclude that,
based on both the ray dynamics calculations and the wave calculations, the 
emission directionality is determined mainly by the structure of the 
unstable manifolds. 

\begin{figure*}
\centering
\includegraphics[clip=true,width=16cm]{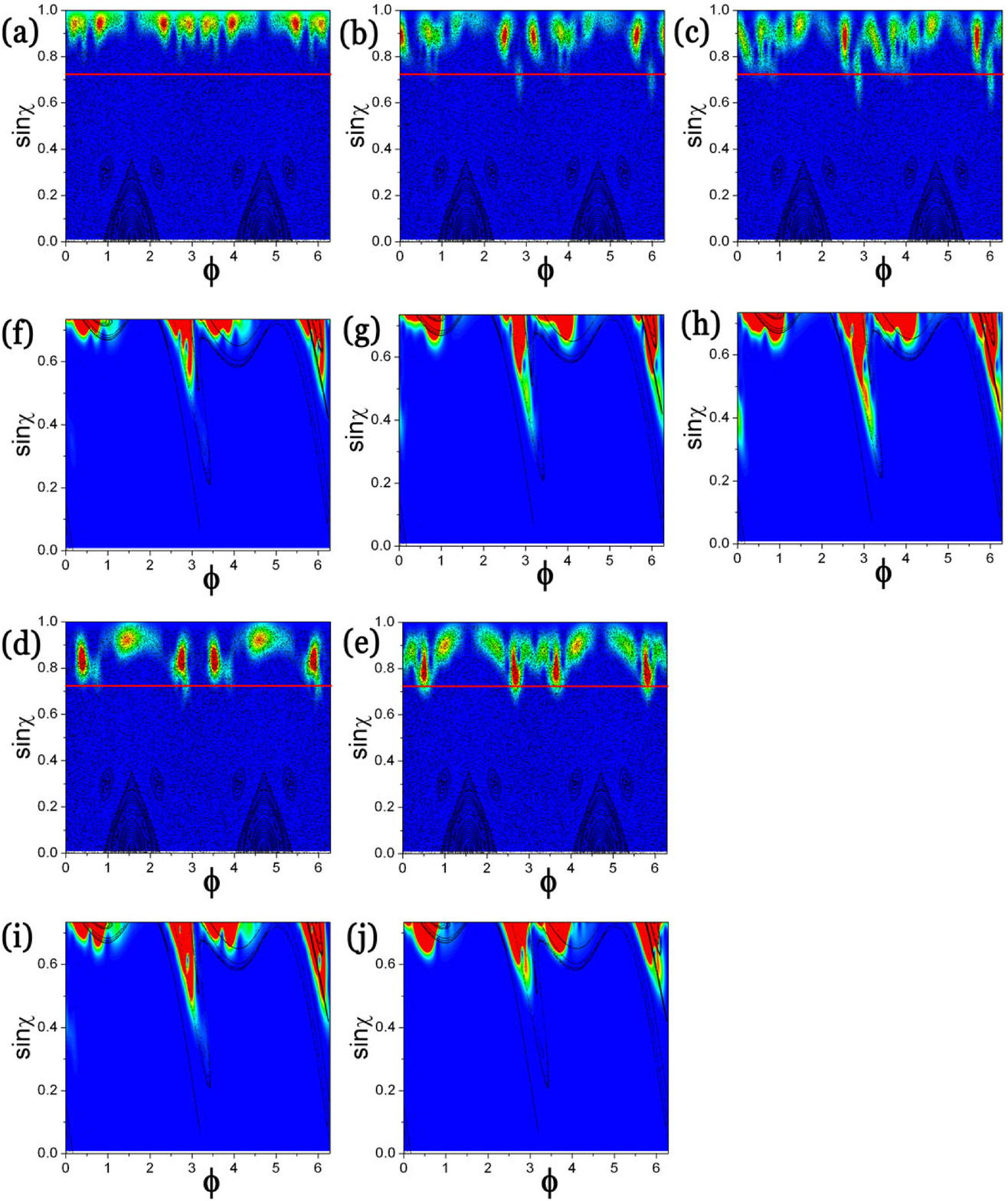}
\caption{Husimi plots of various modes for $\varepsilon=0.16$ whose
complex size parameters are respectively (a) $228.87-i0.002$
($l=1$), (b) $232.27-i0.018$ ($l=2$), (c) $226.30-i0.027$ ($l=3$),
(d) $227.65-i0.018$ ($l=4$), and (e) $230.86-i0.021$ ($l=5$).
(f)-(j) expanded views of the Husimi plots of (a)-(f) for the region below 
the critical angle. The critical line defined by $sin \chi_c = 1/n$
 is indicated with the red horizontal lines on (a)-(e). The unstable manifolds 
are shown in the background with black curves on (f)-(j).} \label{fig:husimi200}
\end{figure*}

\begin{figure}
\centering
\includegraphics[clip=true,width=8.5cm]{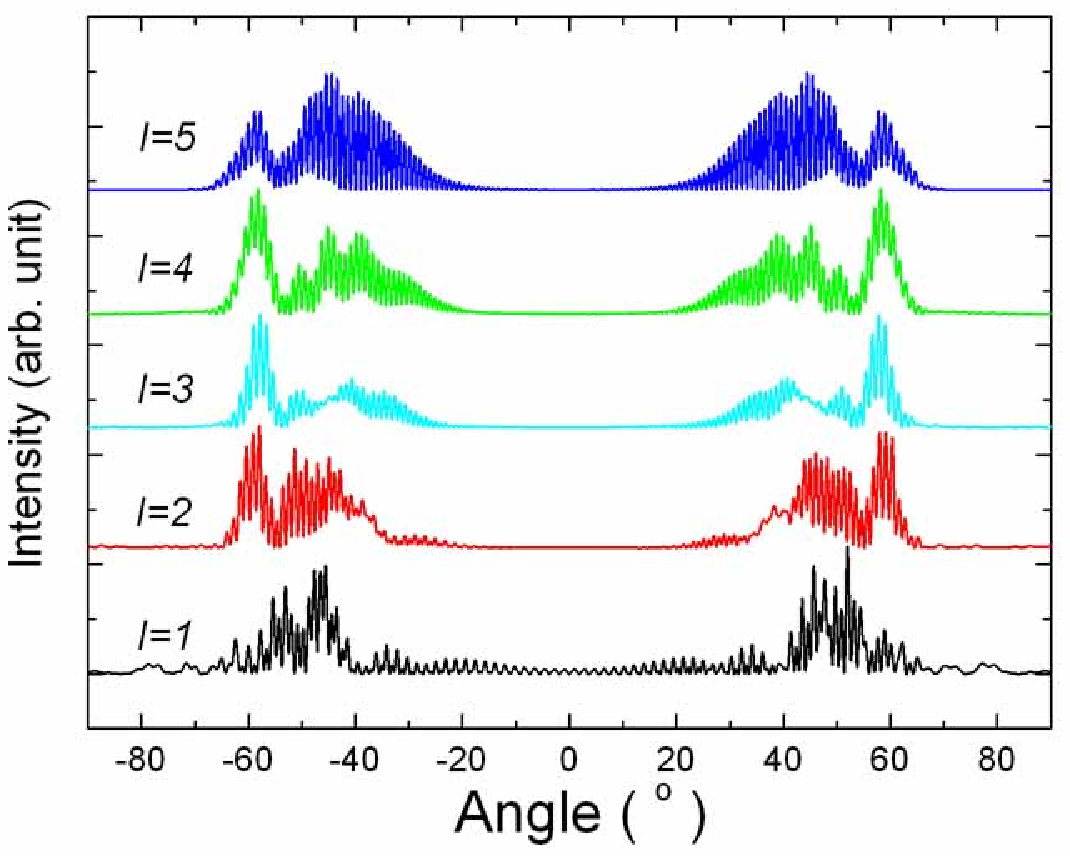}
\caption{The far-field emission distributions associated with the
five modes presented in Fig.~\ref{fig:husimi200}.}
\label{fig:farfield200}
\end{figure}

In order to show that the above conclusion is quite general, we
also investigate the Husimi plots and the far-field emission
directionality of the modes for larger $nkr$, namely $\sim 200$.
Figures \ref{fig:husimi200} (a)-(e) show a variety of shapes of Husimi plots of the modes:
For example, (b) is an octagonal and (d) is a hexagonal scarred mode.
Another interesting point found in Fig. (10) is that each mode does
 not have any
considerable overlap with the others.
In this sense, every mode in Fig.~\ref{fig:husimi200} is quite distinct
from one another. Surprisingly again, the far-field emission
directions are almost identical for all such five {\em distinct}
modes as shown in Fig.~\ref{fig:farfield200}. As mentioned above,
it is also attributed to the geometry of the unstable manifolds as
clearly seen in Figs.~\ref{fig:husimi200}(f)-(j). Therefore, we
call the identical far-field emission directionality determined from
the unstable manifolds a {\em universal} directionality.

\section{Conclusion}
In this work we have studied spectral properties and mode structure of
a quadrupolarly deformed microcavity and the output directionality of the
radiation emitted from it. We summerize the main conclusion of our work.

(1) When the system is open like a QDM, chaotic orbits
no longer easily form a quasieigenmode since their
Q values are expected to be very low due to the refractive
escape. Consequently the localized modes such as
scarred ones are more frequently observed in comparison 
with the usual closed chaotic system. This explains the 
regular structure of the spectrum with well-defined level 
spacings even at large deformations.
 
(2) The direction of the radiation emitted from a quadrupolar microcavity
resonator is determined mainly by the geometry of the unstable manifolds of
the resonator. As the geometry of the unstable manifolds is governed by classical dynamical properties of the resonator and thus is independent of the structure of the excited modes, the output directionality exhibits universality, largely
independent of the degree of deformation and the pump power.

\begin{acknowledgments}
This work was supported by the KOSEF Grant (R01-2005-000-10678-0).
JBS and HWL were supported by a grant from KRISS. SNU group is supported
by the KOSEF Grants(NRL-2005-01371). SWK was supported
 by the KRF Grant (KRF-2004-005-C00044).
\end{acknowledgments}

\bibliography{paper}

\begin{thebibliography}{25}
\expandafter\ifx\csname natexlab\endcsname\relax\def\natexlab#1{#1}\fi
\expandafter\ifx\csname bibnamefont\endcsname\relax
  \def\bibnamefont#1{#1}\fi
\expandafter\ifx\csname bibfnamefont\endcsname\relax
  \def\bibfnamefont#1{#1}\fi
\expandafter\ifx\csname citenamefont\endcsname\relax
  \def\citenamefont#1{#1}\fi
\expandafter\ifx\csname url\endcsname\relax
  \def\url#1{\texttt{#1}}\fi
\expandafter\ifx\csname urlprefix\endcsname\relax\def\urlprefix{URL }\fi
\providecommand{\bibinfo}[2]{#2}
\providecommand{\eprint}[2][]{\url{#2}}

\bibitem[{\citenamefont{Yamamoto and Slusher}(1993)}]{yamamoto}
\bibinfo{author}{\bibfnamefont{Y.}~\bibnamefont{Yamamoto}} \bibnamefont{and}
  \bibinfo{author}{\bibfnamefont{R.~E.} \bibnamefont{Slusher}},
  \bibinfo{journal}{Physics Today} \textbf{\bibinfo{volume}{46}},
  \bibinfo{pages}{66} (\bibinfo{year}{1993}).

\bibitem[{\citenamefont{N\"ockel and Stone}(1997)}]{nockel}
\bibinfo{author}{\bibfnamefont{J.}~\bibnamefont{N\"ockel}} \bibnamefont{and}
  \bibinfo{author}{\bibfnamefont{A.~D.} \bibnamefont{Stone}},
  \bibinfo{journal}{Nature} \textbf{\bibinfo{volume}{385}}, \bibinfo{pages}{45}
  (\bibinfo{year}{1997}).

\bibitem[{\citenamefont{Stone}(2000)}]{stone1}
\bibinfo{author}{\bibfnamefont{A.~D.} \bibnamefont{Stone}},
  \bibinfo{journal}{Physica A} \textbf{\bibinfo{volume}{288}},
  \bibinfo{pages}{130} (\bibinfo{year}{2000}).

\bibitem[{\citenamefont{Gmachl et~al.}(1998)\citenamefont{Gmachl, Capasso,
  Narimanov, N\"ockel, Stone, Faist, Sivco, and A.Cho}}]{gmachl}
\bibinfo{author}{\bibfnamefont{C.}~\bibnamefont{Gmachl}},
  \bibinfo{author}{\bibfnamefont{F.}~\bibnamefont{Capasso}},
  \bibinfo{author}{\bibfnamefont{E.~E.} \bibnamefont{Narimanov}},
  \bibinfo{author}{\bibfnamefont{J.}~\bibnamefont{N\"ockel}},
  \bibinfo{author}{\bibfnamefont{A.~D.} \bibnamefont{Stone}},
  \bibinfo{author}{\bibfnamefont{J.}~\bibnamefont{Faist}},
  \bibinfo{author}{\bibfnamefont{D.}~\bibnamefont{Sivco}}, \bibnamefont{and}
  \bibinfo{author}{\bibnamefont{A.Cho}}, \bibinfo{journal}{Science}
  \textbf{\bibinfo{volume}{280}}, \bibinfo{pages}{1556} (\bibinfo{year}{1998}).

\bibitem[{\citenamefont{Rex et~al.}(2002)\citenamefont{Rex, Tureci, Schwefel,
  Chang, and Stone}}]{rex}
\bibinfo{author}{\bibfnamefont{N.~B.} \bibnamefont{Rex}},
  \bibinfo{author}{\bibfnamefont{H.~E.} \bibnamefont{Tureci}},
  \bibinfo{author}{\bibfnamefont{H.~G.~L.} \bibnamefont{Schwefel}},
  \bibinfo{author}{\bibfnamefont{R.~K.} \bibnamefont{Chang}}, \bibnamefont{and}
  \bibinfo{author}{\bibfnamefont{A.~D.} \bibnamefont{Stone}},
  \bibinfo{journal}{Phys. Rev. Lett.} \textbf{\bibinfo{volume}{88}},
  \bibinfo{pages}{094102} (\bibinfo{year}{2002}).

\bibitem[{\citenamefont{Lee et~al.}(2002)\citenamefont{Lee, Lee, Chang, Moon,
  Kim, and An}}]{lee}
\bibinfo{author}{\bibfnamefont{S.-B.} \bibnamefont{Lee}},
  \bibinfo{author}{\bibfnamefont{J.-H.} \bibnamefont{Lee}},
  \bibinfo{author}{\bibfnamefont{J.-S.} \bibnamefont{Chang}},
  \bibinfo{author}{\bibfnamefont{H.-J.} \bibnamefont{Moon}},
  \bibinfo{author}{\bibfnamefont{S.~W.} \bibnamefont{Kim}}, \bibnamefont{and}
  \bibinfo{author}{\bibfnamefont{K.}~\bibnamefont{An}}, \bibinfo{journal}{Phys.
  Rev. Lett.} \textbf{\bibinfo{volume}{88}}, \bibinfo{pages}{033903}
  (\bibinfo{year}{2002}).

\bibitem[{\citenamefont{Schwefel et~al.}(2004)\citenamefont{Schwefel, Rex,
  Tureci, Chang, Stone, ben Massoud, and Zyss}}]{harald}
\bibinfo{author}{\bibfnamefont{H.~G.~L.} \bibnamefont{Schwefel}},
  \bibinfo{author}{\bibfnamefont{N.~B.} \bibnamefont{Rex}},
  \bibinfo{author}{\bibfnamefont{H.~E.} \bibnamefont{Tureci}},
  \bibinfo{author}{\bibfnamefont{R.~K.} \bibnamefont{Chang}},
  \bibinfo{author}{\bibfnamefont{A.~D.} \bibnamefont{Stone}},
  \bibinfo{author}{\bibfnamefont{T.}~\bibnamefont{ben Massoud}},
  \bibnamefont{and} \bibinfo{author}{\bibfnamefont{J.}~\bibnamefont{Zyss}},
  \bibinfo{journal}{J. Opt. Soc. Am. B} \textbf{\bibinfo{volume}{21}},
  \bibinfo{pages}{923} (\bibinfo{year}{2004}).

\bibitem[{\citenamefont{Kim and Lee}(1998)}]{swkim}
\bibinfo{author}{\bibfnamefont{S.~W.} \bibnamefont{Kim}} \bibnamefont{and}
  \bibinfo{author}{\bibfnamefont{H.~W.} \bibnamefont{Lee}},
  \bibinfo{journal}{Phys. Rev. E} \textbf{\bibinfo{volume}{59}},
  \bibinfo{pages}{5384} (\bibinfo{year}{1998}).

\bibitem[{\citenamefont{Percival}(1973)}]{percival}
\bibinfo{author}{\bibfnamefont{I.~C.} \bibnamefont{Percival}},
  \bibinfo{journal}{Journal of Physics B} \textbf{\bibinfo{volume}{B 6}},
  \bibinfo{pages}{L229} (\bibinfo{year}{1973}).

\bibitem[{\citenamefont{Bohigas et~al.}(1993)\citenamefont{Bohigas, Tomsovic,
  and Ullmo}}]{bohigas}
\bibinfo{author}{\bibfnamefont{O.}~\bibnamefont{Bohigas}},
  \bibinfo{author}{\bibfnamefont{S.}~\bibnamefont{Tomsovic}}, \bibnamefont{and}
  \bibinfo{author}{\bibfnamefont{D.}~\bibnamefont{Ullmo}},
  \bibinfo{journal}{Physics Report} \textbf{\bibinfo{volume}{223}},
  \bibinfo{pages}{43} (\bibinfo{year}{1993}).

\bibitem[{\citenamefont{N\"ockel et~al.}(1994)\citenamefont{N\"ockel, Stone,
  and R.K.Chang}}]{nockel2}
\bibinfo{author}{\bibfnamefont{J.}~\bibnamefont{N\"ockel}},
  \bibinfo{author}{\bibfnamefont{A.~D.} \bibnamefont{Stone}}, \bibnamefont{and}
  \bibinfo{author}{\bibnamefont{R.K.Chang}}, \bibinfo{journal}{Optics Letter}
  \textbf{\bibinfo{volume}{19}}, \bibinfo{pages}{1693} (\bibinfo{year}{1994}).

\bibitem[{\citenamefont{Chang et~al.}(2000)\citenamefont{Chang, N\"ockel,
  Chang, and stone}}]{Chang}
\bibinfo{author}{\bibfnamefont{S.~S.} \bibnamefont{Chang}},
  \bibinfo{author}{\bibfnamefont{J.}~\bibnamefont{N\"ockel}},
  \bibinfo{author}{\bibfnamefont{R.~K.} \bibnamefont{Chang}}, \bibnamefont{and}
  \bibinfo{author}{\bibfnamefont{A.~D.} \bibnamefont{stone}},
  \bibinfo{journal}{JOSA B} \textbf{\bibinfo{volume}{17}},
  \bibinfo{pages}{1828} (\bibinfo{year}{2000}).

\bibitem[{\citenamefont{N\"ockel et~al.}(2002)\citenamefont{N\"ockel, Stone,
  and R.K.Chang}}]{tureci}
\bibinfo{author}{\bibfnamefont{J.}~\bibnamefont{N\"ockel}},
  \bibinfo{author}{\bibfnamefont{A.~D.} \bibnamefont{Stone}}, \bibnamefont{and}
  \bibinfo{author}{\bibnamefont{R.K.Chang}}, \bibinfo{journal}{Optics Express}
  \textbf{\bibinfo{volume}{10}}, \bibinfo{pages}{752} (\bibinfo{year}{2002}).

\bibitem[{\citenamefont{Takehisa~Harayama and Ikeda}(1999)}]{harayama}
\bibinfo{author}{\bibfnamefont{P.~D.} \bibnamefont{Takehisa~Harayama}}
  \bibnamefont{and} \bibinfo{author}{\bibfnamefont{K.~S.} \bibnamefont{Ikeda}},
  \bibinfo{journal}{Phys. Rev. Lett.} \textbf{\bibinfo{volume}{82}},
  \bibinfo{pages}{3803} (\bibinfo{year}{1999}).

\bibitem[{\citenamefont{Lee et~al.}(2005)\citenamefont{Lee, Ryu, Kwon, Rim, and
  Kim}}]{paichai}
\bibinfo{author}{\bibfnamefont{S.-Y.} \bibnamefont{Lee}},
  \bibinfo{author}{\bibfnamefont{J.-W.} \bibnamefont{Ryu}},
  \bibinfo{author}{\bibfnamefont{T.-Y.} \bibnamefont{Kwon}},
  \bibinfo{author}{\bibfnamefont{S.}~\bibnamefont{Rim}}, \bibnamefont{and}
  \bibinfo{author}{\bibfnamefont{C.-M.} \bibnamefont{Kim}},
  \bibinfo{journal}{Phys. Rev. A} \textbf{\bibinfo{volume}{72}},
  \bibinfo{pages}{061801} (\bibinfo{year}{2005}).

\bibitem[{\citenamefont{Haake}(1991)}]{haake}
\bibinfo{author}{\bibfnamefont{F.}~\bibnamefont{Haake}},
  \emph{\bibinfo{title}{Quantum Signatures of Chaos}}
  (\bibinfo{publisher}{Springer-Verlag}, \bibinfo{year}{1991}).

\bibitem[{\citenamefont{Lamb}(1945)}]{lamb}
\bibinfo{author}{\bibfnamefont{H.}~\bibnamefont{Lamb}},
  \emph{\bibinfo{title}{Hydrodynamics}} (\bibinfo{publisher}{Dover,New York},
  \bibinfo{year}{1945}).

\bibitem[{\citenamefont{Birkhoff}(1927)}]{birk}
\bibinfo{author}{\bibfnamefont{G.~D.} \bibnamefont{Birkhoff}},
  \bibinfo{journal}{Acta. Math} \textbf{\bibinfo{volume}{50}},
  \bibinfo{pages}{359} (\bibinfo{year}{1927}).

\bibitem[{\citenamefont{Lichtenberg and Lieberman}(1983)}]{lichtenberg}
\bibinfo{author}{\bibnamefont{Lichtenberg}} \bibnamefont{and}
  \bibinfo{author}{\bibnamefont{Lieberman}}, \emph{\bibinfo{title}{Regular and
  Stochastic Motion}} (\bibinfo{publisher}{Springer-Verlag New York Inc.},
  \bibinfo{year}{1983}).

\bibitem[{\citenamefont{Kagami and Fukai}(1984)}]{kagami}
\bibinfo{author}{\bibfnamefont{S.}~\bibnamefont{Kagami}} \bibnamefont{and}
  \bibinfo{author}{\bibfnamefont{I.}~\bibnamefont{Fukai}},
  \bibinfo{journal}{IEEE Trans. Antennas Prapag.}
  \textbf{\bibinfo{volume}{32}}, \bibinfo{pages}{455} (\bibinfo{year}{1984}).

\bibitem[{\citenamefont{Wiersig}(2003)}]{wiersig}
\bibinfo{author}{\bibfnamefont{J.}~\bibnamefont{Wiersig}},
  \bibinfo{journal}{Journal of Optics A} \textbf{\bibinfo{volume}{5}},
  \bibinfo{pages}{53} (\bibinfo{year}{2003}).

\bibitem[{\citenamefont{Lee}(1995)}]{hwlee}
\bibinfo{author}{\bibfnamefont{H.-W.} \bibnamefont{Lee}},
  \bibinfo{journal}{Physics Report} \textbf{\bibinfo{volume}{259}},
  \bibinfo{pages}{147} (\bibinfo{year}{1995}).

\bibitem[{\citenamefont{B.~Crespi and Chang}(1993)}]{Crespi}
\bibinfo{author}{\bibfnamefont{G.~P.} \bibnamefont{B.~Crespi}}
  \bibnamefont{and} \bibinfo{author}{\bibfnamefont{S.-J.} \bibnamefont{Chang}},
  \bibinfo{journal}{Phys. Rev. E} \textbf{\bibinfo{volume}{47}},
  \bibinfo{pages}{986} (\bibinfo{year}{1993}).

\bibitem[{\citenamefont{Hentschel and Schomerus}(2003)}]{martina}
\bibinfo{author}{\bibfnamefont{M.}~\bibnamefont{Hentschel}} \bibnamefont{and}
  \bibinfo{author}{\bibfnamefont{H.}~\bibnamefont{Schomerus}},
  \bibinfo{journal}{Europhys. Lett.} \textbf{\bibinfo{volume}{62}},
  \bibinfo{pages}{636} (\bibinfo{year}{2003}).

\bibitem[{\citenamefont{N\"ockel}(1997)}]{nockelth}
\bibinfo{author}{\bibfnamefont{J.}~\bibnamefont{N\"ockel}}, Ph.D. thesis,
  \bibinfo{school}{Yale University} (\bibinfo{year}{1997}).

\end{thebibliography}

\end{document}